\documentclass[12pt,letterpaper]{article}
\usepackage[affil-it]{authblk} 
\usepackage{etoolbox}
\usepackage{lmodern}

\makeatletter
\patchcmd{\@maketitle}{\LARGE \@title}{\fontsize{16}{19.2}\selectfont\@title}{}{}
\makeatother

\title{The Logic of Political Survival Revisited:\\ Consequences of Elite Uncertainty Under Authoritarian Rule}
\author{Tamar Zeilberger\footnote{London School of Economics and Political Science (LSE). Contact: t.zeilberger@lse.ac.uk}}
\affil{}
\usepackage{fancyhdr} 
\usepackage{lastpage} 
\usepackage{extramarks} 
\usepackage{graphicx} 
\usepackage{lipsum} 
\usepackage{setspace}
\usepackage[margin=1.0in]{geometry}
\usepackage{tcolorbox}
\usepackage [english]{babel}
\usepackage [autostyle, english = american]{csquotes}
\MakeOuterQuote{"}

\begin{document}
\date{ }
\maketitle
\begin{singlespace}
\noindent \textbf{Abstract:} Existing research has established that autocrats offer concessions to prevent ouster by their inner circle. This paper examines how those concessions are influenced by the relative uncertainty of an autocrat's inner circle about remaining in that favored body. I take as my starting point the formal model of political survival presented in Bueno de Mesquita et. al.'s \textit{The Logic of Political Survival}. I extend the model to account for variation in the relative uncertainty of an autocrat's inner circle. To make the math tractable, I dispense with convention and introduce comparative statics across two models with different formulations of uncertainty. This exercise reveals a set of conditions under which to expect an increase in the concessions offered by an autocrat with implications for development and democracy. Those findings yield a corresponding set of logical corollaries with potential to further our understanding of authoritarian politics, including an unexamined facet of the ``dictator’s dilemma'' (Wintrobe 1990, 1998) and related incentives for members of an inner circle to permit purges or act to destabilize their ranks. The models also identify a source of policy volatility not found outside of autocracies. Taken together, the findings suggest a need for more research on elite uncertainty in autocracies.
\end{singlespace}

\newpage
\section{Introduction}
Across autocracies, coup d'etat occur at a rate that renders them commonplace when compared to other modes of ouster (Tullock 1987, Svolik 2012, Geddes et al. 2018). That long noted empirical regularity is not particularly puzzling. Autocrats typically rely on some elite subset of the population to secure and maintain power. It logically follows that members of this ``inner circle'' is best equipped to remove them from power.

 Past scholarship well establishes that autocrats offer concessions in the form of power and material goods to prevent ouster by their inner circle (see Tullock 1987; Wintrobe 1990, 1998; Bueno de Mesquita 2003; Svolik 2012; Geddes et al. 2018). Despite that proliferation of research, more work needs to be done to understand how autocrats prevent coup d'etat. This paper addresses one dimension that has yet to be sufficiently addressed. Namely, how the concessions offered by autocrats are affected by their inner circles’ relative uncertainty about remaining in that favored body under the status quo and post coup d'etat. 

I take as my starting point the influential model of political survival presented in Bueno de Mesquita et. al.'s \textit{The Logic of Political Survival}. I extend the model to account for variation in the relative uncertainty experienced by autocrats' inner circles about remaining in that favored body under the status quo and post coup d'etat. To make the math tractable, I dispense with convention and introduce comparative statics across two models with different formulations of uncertainty. This exercise allows for a preliminary examination of the consequences of elite uncertainty under a canonical set of assumptions. Since the authors model political survival for all regime types, I also account for the effect of the relative uncertainty experienced by inner circles in democracies. However, the effect diminishes in those cases. The effect is concentrated around cases that represent autocracies and decreases as the inner circle grows. Larger ``inner circles'' are a feature of democracy where power is more dispersed.

I find that autocrats offer more concessions to inner circles when uncertainty about remaining in that favored body is equal under the status quo and post coup d'etat. Concessions decline when the difference in inner circles' relative uncertainty favors the status quo. These findings have some unexpected implications for development and democratization in autocracies. They also yield a set of surprising logical corollaries, revealing an unexamined facet of the ``dictator’s dilemma'' (Wintroble 1990, 1998) and related incentives for members of an inner circle to permit purges or act to destabilize their ranks. An additional finding uncovers a possible source of policy volatility unique to autocracies. Taken together, my findings suggests a need for more research on elite uncertainty in autocracies.

 The paper proceeds as follows. In the next section, I define and discuss the significance of the ``relative uncertainty'' of an inner circle about remaining in that favored body. In section 3, I present two variations of Bueno de Mesquita et al.'s model of political survival that consider the extreme cases of relative uncertainty: asymmetric uncertainty where members of the inner circle are only uncertain about their fates post coup and equal uncertainty where members of the inner circle are equally uncertain about their fates post coup and under the status quo. In section 4, I provide numerical and graphical examples. Finally, in section 5, I discuss the predictions generated by the models and their implications.

\section{Relative Uncertainty}

Prior to introducing the model and my extension, it is necessary to explicitly define the ``relative uncertainty'' of an autocrat's inner circle about remaining in that favored body. All members of the inner circle benefit from concessions granted to them by an autocrat to guarantee their support. These concessions are greater than those received by any other resident of the polity except the autocrat who naturally benefits most from his rule. It is thus in the interest of those in the inner circle to retain membership. However, although they do not want to lose their benefits, they invariably want to gain more. It is at this juncture that the relative uncertainty of those in the inner circle gains significance.

Each member of the inner circle can potentially gain more through one of two means: (1) personally challenging the autocrat for his seat of power or (2) backing a challenger who promises greater concessions. Following Bueno de Mesquita et al.'s model of political survival, I will only consider the second means by which each member of the inner circle can increase her share of benefits. However, should there be interest, finding when a member will or will not personally pose a challenge is a straight forward exercise in calculating her expected utility based on the probability of a successful coup d'etat and her expected utility from continuing to support the incumbent autocrat. We can account for why one member would personally challenge the incumbent while another would not at any given time with the simple intuition that the former is better equipped for such an undertaking and thus has a higher probability of success.

A member of the inner circle will likewise back a challenger if her expected utility under the new order exceeds her expected utility under the current order. In Bueno de Mesquita et al.'s model of political survival, the incumbent has the advantage since it is assumed that his inner circle is fixed. This means that once a member has gained entree into that inner circle, she is certain to remain there. On the other hand, no member of the inner circle is guaranteed membership in a challenger's inner circle. Thus the model offers a simplification of reality where uncertainty is asymmetric and entirely concentrated around the future actions of a challenger. 

The assumption that an incumbent's inner circle is fixed abstracts away one of the defining features of many autocracies: conflict between an autocrat and his inner circle. History is littered with the bodies and careers of former members of autocrats' inner circles who fell out of favor. The mass and at times violent purges of party members orchestrated by Russia's Joseph Stalin in the mid to late 1930s and China's Mao Zedong from the mid 1960s to mid 1970s during the Cultural Revolution are perhaps the most well known. On the heels of Mao, Uganda's Idi Amin famously bloodied and disappeared ministers and members of his regime who disputed his version of the facts and Iraq's Saddam Hussein laid waste to nearly all of his closest supporters. More recently, since taking power in the early 2010s, North Korea's Kim Jong-Un has ordered a spate of executions of family and party members thought to be key policy advisors in his regime and China's Xi Jinping has led an anti-graft campaign that has resulted in the indictment and imprisonment of numerous top party officials. 

These cases illustrate the perils of membership in an autocrat's inner circle. Membership brings the specter of purge which could mean the loss of a job, freedom, health, or life. Accordingly, any model of autocracy is incomplete if it fails to incorporate the fact that the fates of members of the inner circle are not necessarily fixed and are subject to the whims of their autocrat. Indeed, at one extreme, members of the inner circle can conceivably be equally uncertain about their futures under the reigns of an incumbent and a challenger. 

Purges are not the only source of uncertainty for an autocrat's inner circle. The likelihood of retention can also be reduced by competition within the inner circle. Members of Ethiopia's central ruling council, known as the Derg, learned this firsthand in the 1970s (Geddes et al. 2018). Intense struggles between factions led to high turnover on the Derg. Members who lost in those struggles were often jailed or executed and members who won only fleetingly clung to those victories (Geddes et al. 2018). By the end of the decade, nearly all of the chief founders of the Derg were counted among the casualties (Geddes et al. 2018). Further, in the presence of a weak or weakening autocrat, the incumbent advantage derived from uncertainty about remaining in a favored body under a challenger's reign can rapidly disappear. In the span of less than a month, uncertainty about the chances of continued membership in the inner circle of either the incumbent or challenger dramatically swung from one side to the other for supporters of Egypt's Hosni Mubarak. A fact demonstrated by his military-backed ouster after three decades of rule in February 2011.

There is, however, the other extreme captured in Bueno de Mesquita et al.'s original model of political survival. Some autocrats do indeed have remarkably stable inner circles. After all, autocrats are no less heterogeneous than their democratic counterparts. China's Hu Jintao left the Party largely intact during his decade long rule in the 2000s. Since the reunification of Vietnam in 1975, the ruling Communist Party has mirrored post-Mao China with its long periods of stability. Although these periods have also been broken up by sporadic crackdowns on corruption that have unseated numerous top party officials.

In reality, neither of the abovementioned extremes accurately capture the relative uncertainty experienced by an autocrat's inner circle. Uncertainty about continued membership in the inner circle can be nearly equivalent for an incumbent and challenger. However, some asymmetry will always be present. This is due to the fact that any time in office reveals information about an autocrat. Hence even when an autocrat is prone to mass purges, members' estimates of their chances of retention will be more accurate than those under the rule of a challenger. Similarly, although uncertainty about continued membership in the inner circle is asymmetric, it is never asymmetric to the point that members of the inner circle possess perfect confidence that they will be retained by the incumbent. Autocrats possess superior resources compared to any other individual under their rule and need not contend with third party checks to power. These advantages ensure that membership is only as predicable as the autocrat's desires. 

More often than not, the differences between an inner circle's uncertainty about their fates under an incumbent and challenger fall between those two extremes, where they can take on an infinite number of values. Yet despite empirical inaccuracy, the extreme cases of asymmetric and equal uncertainty are useful to consider because they provide us with a way to formally predict the minimum and maximum concessions an autocrat must offer to retain power. In the subsequent section I use them to present two variations of Bueno de Mesquita et al.'s model of political survival. These models allow for an examination of the amount and composition of concessions offered to an autocrat's inner circle with respect to equal and asymmetric uncertainty.

\section{Models of Political Survival} 

\section*{3.1 Overview }

Beuno de Mesquita et al. assume that, all else being equal, those essential to a leader's hold on power and consequently favored are selected based on a leader's idiosyncratic preferences. In the previous sections I refer to this favored body as the ``inner circle''. Bueno de Mesquita et al. refer to them as the ``winning coalition''. Since I merely offer an extension of their model, I adopt that terminology in this section. 

The leader's idiosyncratic preferences over who is selected for the winning coalition are initially unknown. They are revealed by the leader upon assuming power. In the original version of the model where uncertainty is asymmetric, the leader's preferences, and thus the members of the winning coalition, are fixed ad infinitum. For the variation of the model where uncertainty is equal, I relax this assumption so that the leader's preferences are not fixed. 

The leader offers concessions to the winning coalition to maintain power. These concessions are comprised of public goods and private goods. Bueno de Mesquita et al. follow Olson (1965) and define public goods as all goods that are non-excludable and non-rival. In an extension to the model by Bueno de Mesquita \& Smith (2008), the authors provide examples of pubic goods that range from education and infrastructure to civil liberties such as the freedoms of press and assembly. The conflation of material and legal goods in the context of resource allocation is not without flaws. However, I adopt it for the sake of convenience and substantive interest. Specifically, it allows for the distribution of public goods to serve as a rough proxy for democratization. The question of whether this conflation yields accurate predictions is discussed in the subsequent section. Private goods are defined simply as all goods that only benefit the winning coalition. 

Beuno de Mesquita et al. assume that a leader's first priority is political survival. This is a canonical assumption that predates Beuno de Mesquita et al.'s model.\footnote{For an example, see Geddes (1999).} Empirically, its veracity can be demonstrated by cases such as Mexico's Plutarco Calles who crippled the economy to secure and maintain power, and only set the stage for recovery prior to voluntarily leaving office. The leader secondarily prioritizes the maximization of discretionary resources. Discretionary resources are rents hoarded by the leader and spent according to his whims. They enrich a leader and aid his survival. Thus, the leader will always act to efficiently distribute public and private goods.

Following Beuno de Mesquita et al., when making concessions, the leader considers the threat of a challenger. Both the leader and challenger propose a distribution of goods to procure support subject to a budget constraint. The leader and the challenger also propose winning coalitions of some equal size. A challenger's winning coalition must contain at least one member of the leader's winning coalition. If any member of the leader's winning coalition chooses the challenger's proposal, the challenger assumes power. The proposals of that new leader are then implemented and the members of his winning coalition become known. To gain power, the challenger wants to optimize the distribution of goods to convince members of the leader's winning coalition to defect. The incumbent leader survives as long as his offer is at least as valuable as the offer of the challenger. 

The leader's proposal depends on the winning coalition's relative uncertainty about remaining in that favored body. When uncertainty is concentrated around the challenger's preferences, then the winning coalition will discount the challenger's proposal so the leader can concede less. When members of the winning coalition are equally uncertain about the preferences of the leader and challenger, then the leader can concede no less than the challenger. In sum, the two variations of the model consider different formulations of the winning coalition's uncertainty, so that the leader's offer of goods may be somewhat less than equal to the offer of the challenger or must be equal to ensure survival.

\section*{3.2 General Model}

From Bueno de Mesquita et al., both variations of the model consider an infinitely repeated game with common discount factor $\delta$. Polities are generally comprised of four groups: residents, or all who live in a given polity, ($N$); the selectorate, or all residents who might influence the choice of leader, ($S$); the winning coalition, or those who are essential for the leader to maintain power, drawn from the selectorate ($W$); and the leader ($L$); so that $N\geq S\geq W \geq L$ (Bueno de Mesquita et al. 2003, pp. 38). In each period the leader faces threats from a challenger ($C$), who aspires to obtain power but obeys the preexisting rules of the polity. As a result, both the leader and the challenger propose winning coalitions of size $W$ drawn from a potential pool of size $S$. Importantly, in Bueno de Mesquita et al.'s model, the challenger's winning coalition must contain at least one member of the leader's winning coalition. In order to entice supporters to defect from the leader, the challenger proposes an allocation of public goods ($g$) and private goods ($z$) subject to a budget constraint:\begin{equation}R+Nr\phi(g)\geq pg+Wz\end{equation} Where $R$ is non-tax generated income, $Nr\phi(g)$ is tax generated income ($\phi(g)$ is an increasing concave production function), and added together they represent the government's total revenue. $pg+Wz$ is the cost of providing public and private goods so that, put simply, equation (1) states that an offer cannot exceed the government's existing resources. If any member of the leader's winning coalition prefers the challenger's proposal, the incumbent loses power. The proposals of the new leader are then implemented and the members of his winning coalition become known. 

\section*{3.3 Asymmetric Uncertainty}

To gain power, the challenger wants to maximize the distribution of goods to convince members of the leader's coalition to defect. Since the challenger needs a coalition of size $W$, he has a programming problem of $max_{g,z}v(g)+ u(z)$, where $v(g)+ u(z)$ is the value of the receipt of goods ($v$ and $u$ are increasing concave functions and $u(0)=0$), subject to the budget constraint $R+ Nr\phi(g)-pg-Wz\geq 0$. With this program, $z=\frac{R+Nr\phi(g)-pg}{W}$ , and the first order condition is:
\begin{equation} v_g(g)+ \frac{Nr\phi_g(g)-p}{W}u_z\left(\frac{R+ Nr\phi(g)-pg}{W}\right) = 0 .\end{equation} Bueno de Mesquita \& Smith use $\hat{g}$ and $\hat{z}$ to represent the maximum allocations of public and private goods. In the immediate period $v(\hat{g})+ u(\hat{z})$ represents the challenger's best offer. Members of the coalition who defect from the incumbent leader can expect to receive this offer in the immediate period \textit{but} the challenger cannot credibly commit to this offer in the future. This is because it is not until the challenger has obtained power that he reveals the membership of his winning coalition of size $W$ and provides the allocation of public and private goods, $g^*$ and $z^*$, that ensure future survival. Since each member of $S$ has a $\frac{W}{S}$ chance of being a member of his winning coalition, the best a challenger can credibly offer is $v(\hat{g})+ u(\hat{z})+\frac{\delta}{ 1-\delta}v(g^*)+\frac{\delta}{1-\delta}\frac{W}{S}u(z^*).$ By contrast, when uncertainty is asymmetric, the composition of the leader's winning coalition is known and fixed. As a result, the leader retains power as long as the value of his offer $\frac{1}{1-\delta}(v(g^*)+ u(z^*))$ is at least equal to the offer made by the challenger for the current and future periods. Thus, if these offers are equated, to maintain power the leader must satisfy the constraint:
\begin{equation}Select(g,z) = v(g)+ u(z)-v(\hat{g})-u(\hat{z})+ \frac{\delta}{1-\delta}\left(1-\frac{W}{S}\right)u(z) \geq 0\end{equation}
The leader will prefer to select some allocation of goods that maximizes discretionary resources subject to Select(g,z). Accordingly, the leader's programming problem is $max_{g,z}R+ Nr\phi(g)-pg-Wz$ subject to $v(g)+ u(z)-v(\hat{g})-u(\hat{z})+\frac{\delta}{1-\delta}\left(1-\frac{W}{S}\right)u(z) \geq 0.$ The optimal concession of public and private goods implies the first order condition:\footnote{Proof is in the appendix. The assumption, although not the terminology, of asymmetric uncertainty  comes from Bueno de Mesquita et al. (2003) and Bueno de Mesquita \& Smith (2008). However, Bueno de Mesquita \& Smith (2008) do not explicitly derive equation (4) in their paper.}
\begin{equation}
AFOC(g,z)=\frac{(1-\delta)S}{S-\delta W}v_g(g)+\frac{Nr\phi_g(g)- p}{W}u_z(z) = 0
\end{equation}
\section*{3.4 Equal Uncertainty}

$Select(g,z)$ represents that, when uncertainty is asymmetric, the leader's winning coalition will discount the challenger's offer at a higher rate. The leader can accordingly offer less than the challenger and still remain in power. However, if the leader's preferences are not fixed than this constraint is no longer applicable.

When members of the leader's coalition are equally uncertain about the prospects of being retained by the leader and challenger in future periods, the leader's offer must be identical to the challenger's in order to maintain power. This is because  the leader's coalition discounts his and the challenger's offer at the same rate so the leader can do no less than the challenger. The challenger will offer as much as possible in order to procure support so the leader must likewise maximize the distribution of goods to retain members of his coalition. Since the leader must keep a coalition of size $W$, he has a programming problem of $max_{g,z}v(g)+ u(z)$ subject to the budget constraint $R+ Nr\phi(g)-pg-Wz\geq 0$. This program implies the following first order condition:\footnote{Equation (5) here is explicitly provided by Bueno de Mesquita \& Smith (2008, pp. 176) but arrived at differently. For Bueno de Mesquita \& Smith it represents a simplified first order condition, obtained by substituting $\zeta(g)$ for $u(z)$. If interested, see Bueno de Mesquita \& Smith (2008, pp. 176) for details.}\begin{equation} 
EFOC(g,z)=v_g(g)+ \frac{Nr\phi_g(g)-p}{W}u_z(z) = 0 .
\end{equation} 

\section{Results} 

Using numerical examples N=S=10,000, p=200, R=1000, $\delta$=.55, r=.5, and the parameter W=300 provided by Bueno de Mesquita \& Smith (2008), and the likewise provided functions $v(g)=\sqrt{g}$, $u(z)=\sqrt{z}$, $\phi(g)=\sqrt{g}$, the effects of variation in relative uncertainty can be demonstrated. For example, in the model that considers asymmetric uncertainty (using equation 4), the leader allocates 352.06 public goods, 43.35 private goods, and 20779.97 leftover for discretionary resources. In the model that considers equal uncertainty (using equation 5), the leader allocates 602.24 public goods, 51.75 private goods, and has 0 leftover discretionary resources. In both of the models, the leader makes concessions to circumvent threats posed by a challenger. A comparison of these two models demonstrates that the presence of asymmetric uncertainty allows for a lower sufficient allocation of public and private goods by the leader to ensure political survival. It also allows the leader to hoard discretionary resources. In the presence of equal uncertainty, the leader has no such opportunity since their winning coalition will defect to a challenger.

Figure 1 (pp. 13) plots the optimal allocation of public goods to retain power against the size of the winning coalition, when uncertainty is asymmetric $(AFOC(g,z)=0)$ and equal $(EFOC(g,z)=0)$. Comparing the two panels of the figure shows that when the winning coalition experiences asymmetric uncertainty (top panel), the leader will indeed allocate less public goods, although it bears mention that this effect diminishes as the size of the winning coalition grows. Substantively, the diminishing effect can be interpreted to imply that the uncertainty experienced by those necessary for an autocrat to maintain power has a greater effect when their influence is less dispersed. Figure 1 also shows that leaders will rely more on public goods instead of the increasingly expensive private goods as the number of those necessary to maintain power grows. That is supported by Figure 2 (pp. 14) which plots the optimal allocation of private goods to retain power against the size of the winning coalition, when uncertainty is asymmetric and equal. Figure 2 shows the rapidly shrinking allocation of private goods as the size of the winning coalition grows. It also shows that, like public goods, less private goods will be allocated when the winning coalition is more certain about retention under the rule of the incumbent than the rule of the challenger as demonstrated by the above numerical example.\bigskip \bigskip

\begin{center}\textbf{Figure 1.} Optimal Public Good Provision (Vertical Axis) Against Coalition Size (Horizontal Axis), With Asymmetric Uncertainty (top panel) and \newline Equal Uncertainty (bottom panel)\end{center}
\begin{center}
\includegraphics[width=0.45\columnwidth]{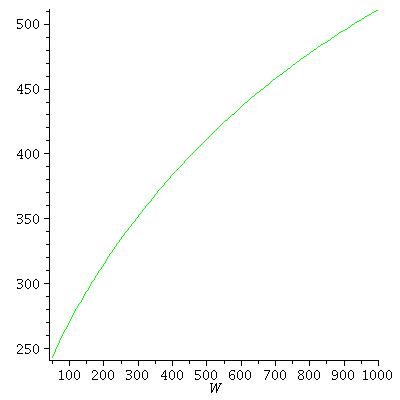}  
\end{center}
\begin{center}
\includegraphics[width=0.45\columnwidth]{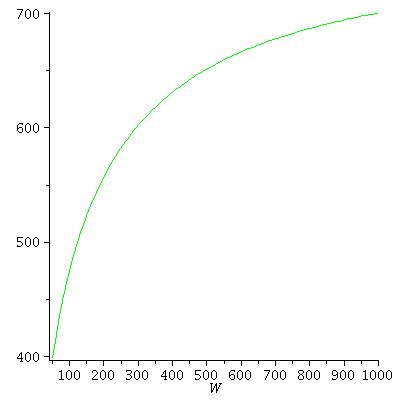}  
\end{center}

\begin{center}
\textbf{Figure 2.} Optimal Private Good Provision (Vertical Axis) Against Coalition Size (Horizontal Axis), With Asymmetric Uncertainty (top panel) and \newline Equal Uncertainty (bottom panel)
\end{center}

\begin{center}
\includegraphics[width=0.45\columnwidth]{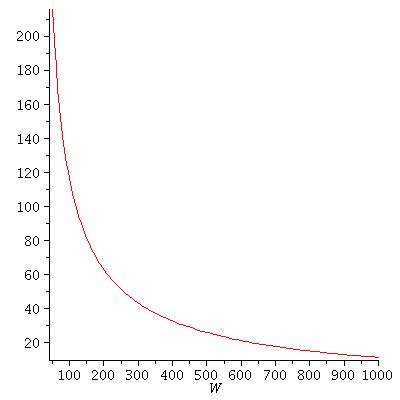}  
\end{center}

\begin{center}
\includegraphics[width=0.45\columnwidth]{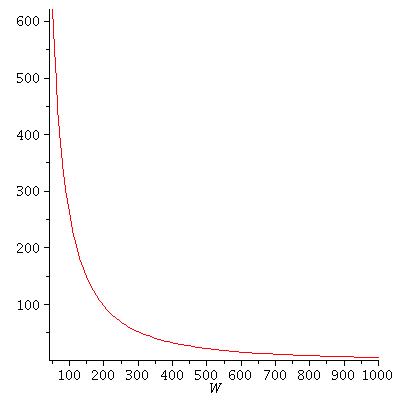}  
\end{center}

\section{Implications} 

The models predict that an autocrat will concede more private goods and public goods when the winning coalition experiences equal uncertainty. The implications of that prediction become clearer when we explicitly consider what constitutes private goods and public goods. Private goods can be direct cash transfers but also activities that produce excludable privileges such as sanctioned rent seeking and market manipulation. Public goods can be large-scale spending projects but, according to Bueno de Mesquita et al., can also include measures of democratization such as the right to public assembly. An increase in either type of good can therefore have far reaching economic and political consequences. With that in mind, it is worth considering whether the predictions from the models conform to known cases and taking note of their outcomes.

That an autocrat will increase private goods when the relative uncertainty experienced by the winning coalition is more equal appears readily apparent in the purge prone autocracies detailed in section 2, which were each riddled with corruption and inefficient policy. However, the same cannot be said about the prediction that an autocrat will increase public goods under the same conditions. This is particularly counterintuitive given that it can imply some greater measure of democratization. One need only recall the rule of Stalin or Amin to find an apparent contradiction. Those discrepancies may support the intuition that the conflation of material and legal public goods is inappropriate here. Public goods also represent a broad class of goods from which an autocrat can pick and choose. Additionally, it is worth noting that for small winning coalitions few public goods are predicted by the models.  It is likewise possible that in some of the cases mentioned previously in this paper, potential challengers were sufficiently weakened so that uncertainty remained more asymmetric. Alternatively, a more encompassing list of cases may actually lend support for the public goods allocation predicted by the models.  

Even if autocrats substitute more private goods with public goods than predicted by the models, an increase in any concessions results in less discretionary resources under their control. Since discretionary resources are linked to political survival (Bueno de Mesquita et al. 2003), the models predict that political survival is more likely for an autocrat whose winning coalition experiences more asymmetric uncertainty. That link reveals a new facet of what is commonly referred to as the dictator’s dilemma (Wintrobe 1990, 1998). The dictator’s dilemma refers to the choice an autocrat makes between the use of repression so that subjects are too fearful to dissent and the ability to identify those who would attempt ouster. To preserve the latter, an autocrat does not solely rely on repression to maintain power. Instead, an autocrat substitutes some repression with concessions to those necessary to maintain power. The findings from the models suggest that repression not only has potential to obscure opposition in an autocracy but also to inflate the concessions necessary to safeguard support. That follows from the stylized fact that repression increases the uncertainty experienced by the winning coalition under the status quo and therefore has the potential to make their relative uncertainty more equal. On the other hand, the absence of repression may embolden a challenger since there is less fear of reprisal, which also poses a threat to political survival. An addendum to the dictator's dilemma emerges from that tradeoff.

The additional cost of repression revealed above also provides an explanation for why members of an autocrat's winning coalition permit purges of their ranks. A member of the winning coalition can be better off when uncertainty under the status quo approaches the uncertainty experienced post coup, so long as the member avoids purge. Naturally, the reason is that an autocrat is compelled to concede more resources when the winning coalition experiences more equal uncertainty. A logical corollary of that implication is that members of an autocrat’s winning coalition have incentive to destabilize their ranks. Thus far, efforts by members of an autocrat’s winning coalition to destabilize their ranks have typically been studied only in the contexts of competition for finite resources, succession, or attempted ouster (Bueno de Mesquita et al. 2003; Padró i Miquel 2007; Acemoglu et al. 2008; North et al. 2009; Boix \& Svolik 2013). Little attention has been paid to efforts to extract more resources from an autocrat who otherwise has sufficient support and will remain in power.

That the effects found in this paper diminish as the size of the winning coalition grows constitutes yet another contribution. A large literature examines the relationship between political institutions and the concessions offered by leaders (McGuire \& Olson 1996; Bueno de Mesquita et al. 2003; Lizzeri \& Persico 2004; Acemoglu \& Robinson 2001, 2006; Deacon 2009). Among those works is the model for which this paper offers an extension. According to their findings, based on economies of scale, democracies are more likely to offer public goods as concessions whereas autocracies are more likely to rely on private goods that can narrowly benefit only those few necessary to maintain power. This paper naturally examines that dynamic and unsurprisingly finds support for that conclusion. However, it also uncovers an additional source of variation between democracies and autocracies. Specifically, it reveals a potential source of policy volatility in autocracies that is not present in democracies since only the former are sensitive to changes in the relative uncertainty experienced by the winning coalition. That same mechanism has potential to generate variation across autocracies. In both cases, it provides a previously unexamined source of variation and an explanation for aberrant cases where autocrats make concessions that do not conform to expectations. Altogether, these findings suggest a need for more research on elite uncertainty in autocracies.

\section*{References}

\noindent Acemoglu, Daron, and James A. Robinson. 2001. ``A Theory of Political Transitions.'' American Economic Review, 91 (4): 938-963.\bigskip

\noindent Acemoglu, Daron and James Robinson. 2005. \textit{Economic Origins of Dictatorship and Democracy}. Cambridge: Cambridge University Press. \bigskip

\noindent  Buchanan, James M., Tullock, Gordon (1962). The Calculus of Consent: Logical Foundations of Constitutional Democracy.\bigskip

\noindent Bueno de Mesquita, Bruce, Alastair Smith, Randolph Siverson, and James Morrow. 2003. 
\textit{The Logic of Political Survival.} Cambridge, MA: MIT Press.\bigskip

\noindent Bueno De Mesquita, Bruce, and Alastair Smith. ``Political Survival and Endogenous
Institutional Change.'' Comparative Political Studies 42.2 (2008): 167-97.\bigskip

\noindent Deacon, Robert T. ``Public good provision under dictatorship and democracy.'' Public Choice 139 (2009): 241–262.\bigskip

\noindent Geddes, Barbara. ``Authoritarian Breakdown:
Empirical Test of a Game Theoretic Argumen'' Presented at annual meeting of the American Political Science Association, Atlanta, September 1999.\bigskip

\noindent Geddes, Barbara; Joseph Wright; and Erica Frantz. 2018. \textit{How Dictatorships Work.} New York: Cambridge University Press.\bigskip

\noindent Knight, Frank. 1921. \textit{Risk, Uncertainty and Profit}. \bigskip

\noindent Lizzeri, Alessandro and Nicola Persico. ``Why did the Elites Extend the Suffrage? Democracy and the Scope of Government, with an Application to Britain's ``Age of Reform'' '', The Quarterly Journal of Economics, Volume 119, Issue 2 (2004): 707–765.\bigskip

\noindent McGuire, Martin C., and Mancur Olson. ``The Economics of Autocracy and Majority Rule: The Invisible Hand and the Use of Force.'' Journal of Economic Literature 34, no. 1 (1996): 72-96.\bigskip

\noindent North, D. C., Wallis, J. J., Weingast, B. R., and Ot´ahal, T. (2009). \textit{Violence and social order: a conceptual framework for interpreting recorded human history}, volume 20. Cambridge
University Press \bigskip

\noindent Olson, Mancur. 1965. \textit{The Logic of Collective Action.} Cambridge, MA: Harvard University Press.\bigskip

\noindent  Padró i Miquel, Gerard. “The Control of Politicians in Divided Societies: The Politics of Fear.” The Review of Economic Studies, vol. 74, no. 4, 2007, pp. 1259–1274.\bigskip

\noindent Svolik, Milan. 2012. \textit{The Politics of Authoritarian Rule}. Cambridge University Press.\bigskip

\noindent Tullock, G. 1987. \textit{Autocracy}. Dordrecht: Nij\bigskip

\noindent Wintrobe, R. 1990. ``The tinpot and the totalitarian: an economic theory of dictatorship.'' American Political Science Review , 84, 849-872. \bigskip

\noindent Wintrobe, R. 1998. \textit{The political economy of dictatorship}. New York: Cambridge University Press.\bigskip

\noindent Wintrobe, R. ``Autocracy and coups d’etat.'' Public Choice 152, 115–130 (2012). 

\section*{Appendix}

Select(g,z) can be simplified since in equilibrium z*=z, getting the constraint
$$v(g)+u(z)-v(\hat{g})-u(\hat{z})+\frac{\delta}{1-\delta}(1-\frac{W}{S})u(z)\geq 0,$$
and
$$1+\frac{\delta}{1-\delta}(1-\frac{W}{S})=\frac{S-\delta W}{(1-\delta)S},$$
so Select(g,z) further simplifies to
$$v(g)-v(\hat{g})-u(\hat{z})+\frac{S-\delta W}{(1-\delta)S}u(z)\geq0.$$
Replace $\geq$0 and set equal to 0 and take derivative w.r.t. g
$$v_g(g)+\frac{S-\delta W}{(1-\delta)S}u_g(z)=0.$$
By the Chain rule, $u_g(z)=u_z(z)z_g(g)$, so we have
$$v_g(g)+\frac{S-\delta W}{(1-\delta)S}u_z(z)z_g(g)=0,$$
so
$$z_g(g)=-\frac{(1-\delta)S}{S-\delta W}\frac{v_g(g)}{u_z(z)}.$$
The discretionary funds are $R+Nr\phi(g)-pg-Wz$. To maximize, differentiate w.r.t. g and set equal to 0
$$Nr\phi_g(g)-p-Wz_g(g)=0.$$
Now plugging-in the expression for $z_g(g)$ above
$$Nr\phi_g(g)-p-W\frac{(1-\delta)S}{S-\delta W}\frac{v_g(g)}{u_z(z)}=0,$$
so
$$\frac{(1-\delta)S}{S-\delta W}v_g(g)+\frac{Nr\phi_g(g)- p}{W}u_z(z) = 0.$$
\end{document}